\documentclass[a4paper,nofootinbib]{revtex4}
\usepackage{epsfig,amsmath,amssymb}
\def\Qz{{\bf Q}}
\def\qj{{\bf q}_j}

\def\gs{\gamma_S}
\def\ee{$e^+e^-$~}
\def\pp{$pp$~}
\def\ppbar{$p\bar p$~}

\def\cc{${\rm c}\bar{\rm c}$~}

\def\d{{\rm d}}
\def\B{\boldmath}

\def\beq{\begin{equation}}
\def\eeq{\end{equation}}
\def\dndy{\left\langle \frac{\d n_j}{\d y} \right\rangle}

\def\B{\boldmath}
\def\lsim{\raise0.3ex\hbox{$<$\kern-0.75em\raise-1.1ex\hbox{$\sim$}}}
\def\gsim{\raise0.3ex\hbox{$>$\kern-0.75em\raise-1.1ex\hbox{$\sim$}}}

\def\beq{\begin{equation}}
\def\eeq{\end{equation}}

\begin{document}

\title{Predictions of hadron abundances in \B\pp collisions at the LHC}

\author{F. Becattini}
\affiliation{Dipartimento di Fisica, Universit\`a di Firenze, and INFN 
Sezione di Firenze, Italy} 
\author{P. Castorina}
\affiliation{Dipartimento di Fisica, Universit{\`a} di Catania, and INFN 
Sezione di Catania, Italy}
\author{A. Milov}
\affiliation{Department of Particle Physics, Weizmann Institute of Science,
Rehovot, Israel}
\author{H. Satz}
\affiliation{Fakult\"at f\"ur Physik, Universit\"at Bielefeld, Germany}

\begin{abstract}
Based on the statistical hadronization model, we obtain quantitative 
predictions for the relative abundances of hadron species in \pp collisions at the 
LHC. By using the parameters of the model determined at $\sqrt s = 200$ GeV,
and extrapolating the overall normalization from \ppbar collisions at the 
SPS and Tevatron, we find that the expected rapidity densities are almost 
grand-canonical. Therefore, at LHC the ratios between different species become 
essentially energy-independent, provided that the hadronization temperature 
$T_H$ and the strangeness suppression factor $\gamma_S$ retain the stable values
observed in the presently explored range of \pp and \ppbar collisions. 
\end{abstract}

\maketitle


Just before the advent of data from the highest energy hadron collider 
of all times, the LHC, we want to ask what, if anything, we can predict 
quantitatively for the forthcoming measurements dealing with the truly
non-perturbative strong interaction regime. The one feature which has
emerged over the years in multihadron production, from \ee annihilation 
to heavy ion collisions, is its statistical nature.
The relative abundances of the different species are predicted with
remarkable precision by an ideal resonance gas model, with a hadronization
temperature converging to about 170 MeV in the limit of high collision
energy \cite{becabiele,becareview,BCMS,BCMS2,kraus2}, and this feature has 
already been used to make predictions for relative hadron abundances 
expected at the LHC in Pb-Pb collisions \cite{pbm,kraus}.

\medskip

The only aspect which distinguishes elementary from nuclear collisions 
is the rate of strangeness production, which in elementary collisions is 
suppressed by a universal factor $\gamma_S \simeq 0.6$, while in heavy 
ion collisions $\gamma_S \to 1$. The origin of $\gamma_S$ has been 
discussed in various approaches; so far, 
there does not seem to exist a satisfactory explanation of its value in elementary 
collisions. We will therefore treat it as a parameter to be determined empirically,
and use the appearent convergence to an energy-dependent value 
$\gamma_s\simeq 0.6$ in \pp interactions \cite{becareview,BCMS} as input 
for our predictions. In this respect, our predictions for \pp collisions
differ from those of ref.~\cite{kraus}, where the extra strangeness suppression
is implemented through the introduction of a strangeness correlation 
volume\footnote{Unlike $\gs$, this mechanism does not suppress hidden strange 
meson production; hence the two approaches give different 
quantitative predictions for the $\phi$ meson yield.}. However, we will consider as 
well $\gamma_S \simeq 1$ as the other ``extreme''. Clearly this issue is the 
most interesting in the analysis of the forthcoming \pp data.    

\medskip

The statistical model to be used is described in detail in ref.~\cite{becareview}.
We repeat here only essential points and caveats for its specific application
to LHC experiments. The simple analytical formulae for multiplicities derived
within the statistical model (see e.g.\ \cite{BCMS2}) apply in principle to
full phase space multiplicities, since possible charge-momentum correlations
are integrated out. Therefore, in order to apply the same formulae to
midrapidity data, one has to assume that particle ratios there are essentially 
the same as those in full phase space. While such an assumption is certainly
not tenable at low collision energy, it is expected to become valid in sufficiently
high energy collisions with large rapidity coverage. We thus assume that the
primary rapidity density of each species in \pp collisions is given by 
(see e.g. \cite{BCMS,becaheinz}):
\beq\label{form2}
 \dndy^{\rm primary}_{y=0} = 
 \frac{A V T (2J_j+1)}{2\pi^2} 
 \sum_{n=1}^\infty \gs^{N_s n}(\mp 1)^{n+1}\;\frac{m_j^2}{n}\;
 {\rm K}_2\left(\frac{n m_j}{T}\right)\, \frac{Z(\Qz-n\qj)}{Z(\Qz)},
\eeq
where $A$ is a common normalization factor taking into account the ratio of production 
in the mid-rapidity interval to the overall rate; $V$ is a volume (see 
discussion below), $T$ is the temperature, $Z(\Qz)$ is the canonical partition 
function depending on the initial abelian charges $\Qz = (Q,N,S,C,B)$, i.e., 
electric charge, baryon number, strangeness, charm and beauty, respectively; 
$m_j$ and $J_j$ are the mass and the spin of the hadron $j$, and 
$\qj = (Q_j,N_j,S_j,C_j,B_j)$ its corresponding charges; $\gs$ is the extra 
phenomenological factor implementing a suppression of hadrons containing 
$N_s$ strange valence quarks. In the formula (\ref{form2}), the upper 
sign applies to bosons and the lower sign to fermions. For temperature 
values of 160 MeV or higher, Boltzmann statistics corresponding to the 
term $n=1$ only in the series (\ref{form2}) is a very good approximation for 
all hadrons (within 1.5\%) but pions. For resonances, the formula (\ref{form2}) 
is folded with a relativistic Breit-Wigner distribution of the mass $m_j$. 
To the above primary production 
one has to add the secondary production due to the strong and electromagnetic decay 
chains. In our calculation, we include all known resonances up to mass of 1.8
GeV, as well as baryon resonances of $\Lambda$-, $\Delta$- and $\Xi$-type
between 1.8 and 1.92 GeV. Also, it is assumed that these decays do not distort 
noticeably the rapidity distributions. 

\medskip

In formula (\ref{form2}), the volume $V$ appears both as an overall multiplicative 
factor and in the chemical factors $Z(\Qz-n\qj)/Z(\Qz)$ (related to the so-called 
canonical suppression phenomenon). It thus contributes to the determination of
the {\em ratios} of different particle species and cannot be absorbed into an overall 
normalization factor $AV$. We note that this volume, determined by fitting
measured rapidity densities to the formula (\ref{form2}), has no direct
physical meaning. In fact, it corresponds to the volume a cluster would have
if its fully integrated hadronic multiplicities were proportional to the
measured midrapidity densities. Only if one uses experimental $4\pi$
multiplicities does the fitted volume have a more direct physical meaning; in 
the global cluster scheme \cite{becareview}, it gives the sum of the volumes
of the actually produced clusters, for pointlike hadrons. 

\medskip 

The volume factor determined at $\sqrt s = 200$~GeV is $VT^3 = 135 \pm 60$ 
\cite{BCMS2}; it is obtained from an analysis of midrapidity densities
measured at RHIC; hence the large error. In order to extrapolate it to LHC
energy, we assume
that its value evolves with energy the same way as midrapidity densities. 
This requires that $T$ stays constant, for which we have strong evidence, 
as already mentioned. By using five $\d n/d \eta$ values measured for charged 
particles at SPS and Tevatron at energies $200,~546,~630,~900$ and
1800 GeV in \ppbar collisions \cite{data}, we found that this evolution 
is best fitted with a polynomial in $\log \sqrt s$:
\beq\label{poly}
   \frac{\d n_{\rm ch}}{\d \eta}= 
   1.35+0.0375 \log \sqrt s +0.00962 \log^2\sqrt s + 0.00434 \log^3\sqrt s,  
\eeq
with $\sqrt s$ in GeV. With the above coefficients, and renormalizing the right 
hand side so as to obtain $VT^3=135$ at $\sqrt s =200$~GeV, we are able to
estimate the $VT^3$ parameter at larger energies. For instance, at $\sqrt s
= 10$ TeV, the predicted value becomes $VT^3 = 323$. 

\medskip

Here a comment is in order. The parameter $VT^3$ was determined with a large
error in the statistical model analysis of the RHIC data, because it is
strongly anticorrelated to the temperature. Both $T$ and $VT^3$ contribute to 
determine the canonical weight factors in formula (\ref{form2}), but the final 
uncertainty of the weights, as determined from fit errors, is much smaller 
than that of $VT^3$, because of its anticorrelation to $T$. Moreover, the 
canonical weights have another important feature: for large volumes and fixed, 
finite charges, they saturate to their grand-canonical limit 1, so that the 
relevant uncertainties naturally decrease as the energy, and hence the system
size, is increased. This is precisely the case for LHC, where, at least for 
midrapidity densities, the grand-canonical limit seems to be almost attained, 
as we will shortly see. 

\medskip 

In table~\ref{first}, we provide predictions for the ratios between midrapidity
densities of several species and that of charged particles for a temperature
value $T=170$ MeV, using the extrapolated value of $VT^3 (=323)$ parameter at 
$\sqrt s = 10$ TeV, and $\gs=0.6$ and $\gs=1$. It can be seen that the
difference between particle and antiparticle yields is small and not larger 
than 10\%. This is a manifestation of the proximity of the chemical factors
(with special regard to baryon number) to 
their asymptotic value 1 and it implies that the numbers in table~\ref{first} 
are stable against a variation of centre-of-mass energy within the typical LHC 
range, from 1 TeV onwards. Therefore, in this energy region, the main 
source of error on model predictions is the uncertainty on the parameters $T$ and 
$\gs$, whose values are an educated guess based on those determined at $\sqrt s = 200$ 
GeV \cite{BCMS2} and the very mild increasing trend observed for $\gs$ \cite{becareview}. 
The uncertainties can be reasonably estimated to be of the order of 3\% for the 
temperature and 8\% for $\gs$ which are reflected into an error on the ratios 
quoted in the left column of table~\ref{first} depending on particle species,
ranging from few percents for pions up to 20\% for $\phi$ and 40\% for $\Omega$,
which is the worst case.

\begin{table}[!h]\begin{center}
\begin{tabular}{|c|c|c|}
\hline
  Particle         & $(dn/dy)/(dn/dy_{\rm ch})$  & $(dn/dy)/(dn/dy_{\rm ch})$ \\
                   & \B$\gs=0.6$ &\B$\gs=1$   \\			       
\hline

$\pi^0$            & 0.463	 & 0.442     \\ 			      
$\pi^+$            & 0.415	 & 0.392     \\ 			      
$\pi^-$            & 0.412       & 0.389     \\
$K^+$              & 0.0483      & 0.0703    \\
$K^-$              & 0.0474      & 0.0691    \\
$K_S^0$            & 0.0471	 & 0.0681    \\			      
$\eta$             & 0.0499      & 0.0526    \\
$\rho^0$           & 0.0565	 & 0.0508    \\
$\rho^+$           & 0.0561      & 0.0500    \\
$\rho^-$           & 0.0555	 & 0.0496    \\ 	     
$\omega$           & 0.0508      & 0.0449    \\
$\eta'$            & 0.0497	 & 0.0457    \\
$\phi$             & 0.00379	 & 0.00908   \\ 			  
$p$                & 0.0334      & 0.0294    \\
$\bar p$           & 0.0303	 & 0.0271    \\
$\Lambda$          & 0.0115	 & 0.0165    \\ 
$\bar\Lambda$      & 0.0107	 & 0.0156    \\ 			  
$\Xi^-$            & 0.00104     & 0.00254   \\
$\bar\Xi^+$        & 0.000995	 & 0.00245   \\ 			  
$\Omega^-$         & 0.000115    & 0.000474  \\
$\bar\Omega^+$     & 0.000111	 & 0.000464  \\
\hline
\end{tabular}\end{center}
\caption{Predictions of the midrapidity density of hadrons relative to that 
of all charged hadrons at $\sqrt s= 10$ TeV, using an extrapolated energy
dependence and assuming a hadronization temperature $T=170$ MeV. The quoted 
rates {\em do not} include weak decay products.}
\label{first}
\end{table}

\medskip

In table~\ref{second}, we provide the same set of predictions for a temperature
value $T=170$ MeV and $\gs=0.6$ and $\gs=1$, but in the fully grand-canonical
formalism, i.e. for the infinite energy limit. It can be seen that the difference
with respect to previous case is in most cases very small, and it is also well
within the estimated theoretical uncertainty of our main calculation shown in
table~\ref{first}. This indicates that the
LHC is expected to provide hadron abundances corresponding almost to the infinite
energy limit. It is perhaps worthwhile to emphasize this point in more detail.
Hadronization of strongly-interacting system does not depend on its initial 
energy density, and hence not on the initial collision energy. Thus the validity 
of the given predictions for relative abundances does not depend on the functional 
energy dependence of the overall hadron multiplicity. At lower collision energy, 
in the statistical hadronization model, an energy-dependence of relative abundances 
enters through the conservation laws of inner charges (and possible variation 
of $\gs$). When these saturate to the grand-canonical limit at high energies,
the predictions of relative abundances are those of asymptotically stable
thermodynamics. 
  
\begin{table}[!h]\begin{center}
\begin{tabular}{|c|c|c|}
\hline
  Particle         & $(dn/dy)/(dn/dy_{\rm ch})$  & $(dn/dy)/(dn/dy_{\rm ch})$ \\
                   & \B$\gs=0.6$ &\B$\gs=1$   \\			       
\hline
$\pi^0$                 & 0.462      & 0.441	 \\				      
$\pi^+=\pi^-$           & 0.413      & 0.390	 \\				      
$K^+=K^-$               & 0.0480     & 0.0698	 \\
$K_S^0$                 & 0.0473     & 0.0682	 \\				      
$\eta$                  & 0.0497     & 0.0524	 \\
$\rho^0$                & 0.0563     & 0.0506	 \\
$\rho^+=\rho^-$         & 0.0557     & 0.0497	 \\	
$\omega$                & 0.0505     & 0.0477	 \\	
$\eta'$                 & 0.00375    & 0.00455   \\
$\phi$                  & 0.00377    & 0.00903   \\				  
$p=\bar p$              & 0.0321     & 0.0285	 \\
$\Lambda = \bar\Lambda$ & 0.0112     & 0.0162	 \\				  
$\Xi^+=\Xi^-$           & 0.00105    & 0.00254   \\				  
$\Omega^+=\Omega^-$     & 0.000121   & 0.000488  \\
\hline
\end{tabular}\end{center}
\caption{Predictions of the midrapidity density of hadrons relative to that 
of all charged hadrons at $\sqrt s= 10$ TeV in the grand-canonical limit. 
The temperature value is assumed as $T=170$ MeV, and 
the numbers {\em do not} include weak decay products.}
\label{second}
\end{table}

\medskip

Finally, since at the LHC the production cross-section of \cc pairs is 
predicted to be of the order of few mb, we have checked the stability of
the above predictions against the introduction of heavy flavoured hadron
contribution. We have then estimated the same ratios by assuming 30\% 
of events with \cc production and found that the differences with respect
to the no-\cc case are of the order of a percent or less.

\vskip1cm

\section*{References}

\end{document}